\documentclass[a4paper]{jpconf}
\usepackage{latexsym,graphicx}
\begin{document}
\title{Quasinormal modes and thermodynamical aspects of the 3D Lifshitz black hole}

\author{B Cuadros-Melgar$^1$, J de Oliveira$^2$ and C E Pellicer$^3$}

\address{$^1$Physics Department, University of Buenos Aires, 
FCEN-UBA and IFIBA-CONICET, Ciudad Universitaria, Pabell\'on 1, 1428, Buenos Aires, Argentina}
\address{$^2$Instituto de F\'\i sica, Universidade Federal do Mato Grosso, 
78060-900, Cuiab\'a, Brazil}
\address{$^3$Instituto de F\'\i sica, Universidade de S\~ao Paulo, C.P.66318, CEP 05315-970, S\~ao Paulo, Brazil}

\ead{berthaki@gmail.com, jeferson@fisica.ufmt.br, carlosep@fma.if.usp.br}

\begin{abstract}
We consider scalar and spinorial perturbations on a background described by a $z=3$ three-dimensional Lifshitz black hole. We obtained the corresponding quasinormal modes which perfectly agree with the analytical result for the quasinormal frequency in the scalar case. The numerical results for the spinorial perturbations reinforce our conclusion on the stability of the model under these perturbations. We also calculate the area spectrum, which prove to be equally spaced, as an application of our results.
\end{abstract}

\section{Introduction}

New Massive Gravity (NMG) is a novel parity-preserving, unitary~\cite{unit}, power-counting super-renormalizable~\cite{ren}, three-dimensional model describing the propagation of a massive positive-energy graviton with two polarization states of helicity $\pm 2$ in a Minkowski vacuum, whose linearized version is equivalent to the Pauli-Fierz theory for a massive spin-2 field in three dimensions.
The action of NMG consists of a ``wrong sign'' Einstein-Hilbert term plus a quadratic curvature term given by a precise combination of the Ricci tensor and the curvature scalar, which introduces a mass parameter~\cite{nmg}.
As with other massive gravity theories, NMG also admits black hole-type solutions with several asymptotics and additional parameters~\cite{nmg2,bhnmg}. Even though this last feature could challenge the usual Einstein-Hilbert gravity, it is seen that the definition of mass in this new type of black holes is a conserved charge computed from a combination of the black hole parameters, which satisfies the first law of thermodynamics. 

In this contribution we present the results of a study on the stability of the three-dimensional Lifshitz black hole found in the context of the so-called new massive gravity (NMG)~\cite{aggh}. These solutions exhibit the anisotropic scale invariance, $t\rightarrow \lambda^z t$, $\vec x \rightarrow \lambda\vec x$, where $z$ is the dynamical critical exponent. Specifically, we deal with the solutions found for the particular case of $z=3$ and a precise value of the mass parameter. 

We aimed to determine the stability of this kind of black holes by calculating the quasinormal modes (QNM) and quasinormal frequencies (QNF) of scalar and spinorial matter fields in the probe limit, {\it{i.e}}, there are no backreaction effects upon the asymptotically Lifshitz black hole metric. QNM do not depend of the particular initial perturbation that excited the black holes, while QNF depend only on the parameters of the black hole. For this reason QNF are considered the ``footprints'' of this structure. The study of QNM has motivated the development of numerical and analytical techniques for their computation (\cite{kok,nol,bcs}). QNM are important in astrophysical grounds as a way of possible detection of black holes through the observation of its gravitational wave spectrum. They are also relevant in the context of gauge/gravity correspondence~\cite{kachru,alt}, where the Hawking temperature of the black hole is related to the temperature of a thermal field theory defined at the boundary, and QNF can be related to the dual field theory relaxation time. In this work we focus in the application of QNM as probes of stability. Moreover, the QNF we obtain are useful to calculate the area spectrum of Lifshitz black holes. Actually, according to Bekenstein the horizon area of a black hole must be quantized~\cite{bek}, so that the area spectrum has the form $A_n = \gamma n \hbar$, with $\gamma$ a dimensionless constant to be determined. In this way, we find this spectrum by using an approach suggested by Maggiore~\cite{maggiore}, who claims that QNM should be described as damped harmonic oscillators, and the module of the entire QNF is the proper physical frequency that should be used in the Bohr-Sommerfeld quantization of the adiabatic invariant $I=\int dE/\omega(E)$.

In the following sections we describe the model we dealt with, the equations for the scalar and spinorial perturbations, the numerical analysis, and the calculation of the area spectrum. In the last section we discuss our results and conclude.

\section{The black hole solution}

\begin{figure}[htb!]
\begin{center}
\includegraphics[height=8cm]{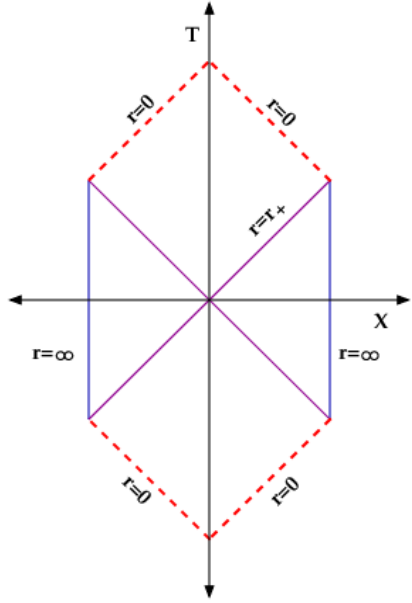}
\end{center}
\caption{Penrose-Carter diagram for the Lifshitz black hole. The singularity at $r=0$ is light-like and covered by a regular event horizon $r_+$.}
\label{penrose}
\end{figure}

The black hole we study along this paper is a solution of the NMG $(2+1)$-dimensional action, 
\begin{equation}\label{action}
S=\frac{1}{16\pi G} \int d^3 x \sqrt{-g} \,\left[R-2\lambda-\frac{1}{m^2}\left(R_{\mu\nu}R^{\mu\nu}-\frac{3}{8}R^2\right)\right]\,,
\end{equation}
where $m$ is the so-called ``relative'' mass parameter, and $\lambda$ is the
three-dimensional cosmological constant.
This theory has a black hole solution with Lifshitz scaling $z=3$ given by
\begin{equation}\label{metric}
ds^{2}=-a(r)\frac{\Delta}{r^{2}}dt^{2}+\frac{r^{2}}{\Delta}dr^{2}+r^{2}d\phi^{2}\,,
\end{equation}
where 
\begin{equation}\label{a}
a(r)=\frac{r^{4}}{l^{4}}\,,
\end{equation}
and
\begin{equation}\label{delta}
\Delta=-Mr^{2}+\frac{r^{4}}{l^{2}}\,,
\end{equation}
with $M$ an integration constant and $l^{2}=-\frac{13}{2\lambda}=-\frac{1}{2m^2}$. Also, the NMG admits as a solution, the well-known Ba\~nados-Teitelboim-Zanelli (BTZ) black hole with the dynamical critical exponent $z=1$. 
The Kretschmann scalar for the metric (\ref{metric}),
\begin{equation}\label{kresh}
R_{\mu\nu\lambda\sigma}R^{\mu\nu\lambda\sigma}= -\frac{4}{l^{4}r^{4}}\left[8r_{+}^{4}-48r_{+}^{2}r^{2}+91r^{4}\right]\,,
\end{equation}
diverges for $r\rightarrow 0$. Thus, the black hole solution has a genuine spacetime singularity at the origin. From the Penrose-Carter diagram in Fig.\ref{penrose} we see that the spacetime singularity at $r=0$ is light-like covered by a regular event horizon at $r=r_{+}=l\sqrt{M}$.

\section{Scalar perturbation equation}

In order to consider the response of the Lifshitz black hole background to a scalar perturbation we consider a scalar field obeying the Klein-Gordon equation,
\begin{equation}\label{kg}
\Box \Phi = \frac{1}{\sqrt{-g}} \partial_M\left(\sqrt{-g}g^{MN} \partial_N \right)\Phi = m^2 \Phi \,,
\end{equation}
where $m$ is the mass of the field $\Phi(t,r,\phi)$.
After decomposing the angular part of the field this equation takes the following form,
\begin{eqnarray}\label{kg3}
-\partial_t ^2 \Psi + \frac{r^4}{l^6} \left(1-\frac{Ml^2}{r^2}\right) \left(\frac{5r^3}{l^2}-3Mr \right) \partial_r \Psi + \frac{r^8}{l^8} \left( 1-\frac{Ml^2}{r^2}\right)^2 \partial_r ^2 \Psi - &&\nonumber \\
-\frac{r^4}{l^6}\left( m^2 r^2 + \kappa^2 \right)\left( 1-\frac{Ml^2}{r^2}\right) \Psi &=&0\,.\phantom{xx}
\end{eqnarray} 

Although this equation has an analytical solution, which we will show later, it is a useful test for our numerical code to check the numerical results. With this goal we need to rewrite (\ref{kg3}) in a Shr\"odinger-like form, 
\begin{equation}\label{schrod-sc}
-\partial_t ^2 X + \partial_{r_*} ^2 X = V(r)X \,,
\end{equation}
where $r_*$ is the so-called {\it tortoise} coordinate given by
\begin{equation}\label{tortoise}
r_* = l^4 \left[ -\frac{1}{M^{3/2}l^3}\; \hbox{arccoth} \left(\frac{r}{l\sqrt{M}}\right) + \frac{1}{Ml^2 r}\right] \,,
\end{equation}
and the scalar effective potential can be written as
\begin{equation}\label{sc-pot}
V(r) = \left(\frac{7}{4l^8} + \frac{m^2}{l^6} \right) r^6 - \left(\frac{5M}{2l^6} + \frac{Mm^2}{l^4} - \frac{\kappa^2}{l^6}\right) r^4 + \left( \frac{3M^2}{4l^4} - \frac{M\kappa^2}{l^4}\right) r^2\,.
\end{equation}

Now let us return to Eq.(\ref{kg3}) and discuss its analytical solution. First of all we set the time dependence of the scalar field as $\Psi(t, r)=R(r)e^{-i\omega t}$, and then redefine the radial coordinate as $r=l\sqrt{M}/y$. The resultant equation has the following solution in terms of the Heun confluent functions,
\begin{eqnarray}\label{solR}
R(y) &=& C_1 y^{2+\alpha} (1-y^2)^{\beta/2}\, \hbox{HeunC}\left(0,\alpha,\beta,-\frac{\beta^2}{4},\frac{\alpha^2}{4}+\frac{\kappa^2}{4M},y^2\right) + \nonumber \\
&&+C_2 y^{2-\alpha} (1-y^2)^{\beta/2}\, \hbox{HeunC}\left(0,-\alpha,\beta,-\frac{\beta^2}{4},\frac{\alpha^2}{4}+\frac{\kappa^2}{4M},y^2\right)\,,\;
\end{eqnarray}
where $C_1$ and $C_2$ are integration constants, while $\alpha=\sqrt{4+m^2 l^2}$, and $\beta=-i\, l\omega/M^{3/2}$. After imposing Dirichlet condition at infinity and the boundary condition of ingoing waves at the horizon we obtain the quasinormal frequency spectrum given by
\begin{equation}\label{freq_analitico}
\omega = 2i \frac{M^{3/2}}{l} \left[ 1+2N+\sqrt{4+m^2 l^2} - \sqrt{7+\frac{3}{2}m^2 l^2 +\frac{\kappa^2}{2M} + (3+6N)\sqrt{4+m^2 l^2}+6N(N+1)}\right]\,,
\end{equation}
where $N$ is a positive integer. These same frequencies were also obtained in subsequent works\cite{luna,sam} by using equivalent analytical procedures.
The imaginary part of the fundamental frequency ($N=0$) is negative provided that
\begin{equation}\label{condition}
\sqrt{7+\frac{3}{2} m^2 l^2 + \frac{\kappa^2}{2M}+3\sqrt{4+m^2 l^2}} > 1+\sqrt{4+m^2 l^2} \,.
\end{equation}
While the asymptotic frequency ($N\rightarrow \infty$) is given by
\begin{equation}\label{asymptotic}
\omega_\infty = -2(\sqrt{6}-2)\,i\, \frac{M^{3/2}}{l} N <0\,.
\end{equation}
Thus, since the imaginary part of the quasinormal frequencies is negative provided that the parameters respect the relation (\ref{condition}), we can conclude that the model is stable under scalar perturbations.

\section{Spinorial perturbation equation}

Let us consider a spinorial field perturbing the Lifshitz black hole background, which obeys the covariant Dirac equation,
\begin{equation}\label{dirac}
i\gamma^{(a)}e_{(a)}^{\phantom{(a)}\mu}\nabla_{\mu}\Psi-\mu_{s}\Psi=0\,,
\end{equation}
where Greek indices refer to spacetime coordinates $(t,r,\phi)$, and
the Latin indices enclosed in parentheses describe the flat tangent 
space in which the triad basis $e_{(a)}^{\phantom{(a)}\mu}$ is defined. 
The spinor covariant derivative $\nabla_{\mu}$ is given by 
\begin{equation}\label{covariant_dev}
\nabla_{\mu}=\partial_{\mu}+\frac{1}{8}\omega_{\mu}^{\phantom{\mu}(a)(b)}\left[\gamma_{(a)},\gamma_{(b)}\right]\,,
\end{equation}
where $\omega_{\mu}^{\phantom{\mu}(a)(b)}$ is the spin connection,
\begin{equation}\label{connection}
 \omega_{\mu}^{\phantom{\mu}(a)(b)}=e_{\nu}^{\phantom{\mu}(a)}\partial_{\mu}e^{(b)\nu}+e_{\nu}^{\phantom{\nu}(a)}\Gamma^{\nu}_{\phantom{\nu}\mu\sigma}e^{\sigma
 (b)}\,,
\end{equation}
$\Gamma^{\nu}_{\phantom{\nu}\mu\sigma}$ are the metric connections, and the gamma matrices we employ along this work are $\gamma^{(0)}=i\sigma_{2}$, $\gamma^{(1)}=\sigma_{1}$, and $\gamma^{(2)}=\sigma_{3}$. 

In order to solve the Dirac equation in a numerical way once again we should adapt it to a Shr\"odinger-like form. First, we redefine the two component spinor as, 
\begin{eqnarray}\label{two-spinor}
\Psi=\left( \begin{array}{c}
\Psi_{1}(t,r,\phi)  \\
\Psi_{2}(t,r,\phi)  \\ \end{array} \right)\,,
\end{eqnarray}
as
\begin{eqnarray}\label{redef-fields}
\Psi_{1}&=&i\left[a(r)\Delta\right]^{1/4}e^{-i\omega t+im\phi}e^{i\theta/2}R_{+}(r),\nonumber \\
\Psi_{2}&=&\left[a(r)\Delta\right]^{1/4}e^{-i\omega t+im\phi}e^{-i\theta/2}R_{-}(r)\,,
\end{eqnarray}
where 
\begin{equation}\label{theta}
\theta=\arctan(\frac{\mu_{s}r}{\hat{m}})\,, \quad m=i\hat m\,.
\end{equation}
And in addition we make $X_{\pm}=R_{+}\pm R_{-}$ to finally have
\begin{equation}\label{dirac-final}
\left(\partial^{2}_{\hat{r}_{*}}+\omega^{2}\right)X_{\pm}=V_{\pm}X_{\pm}\,,
\end{equation}
where $V_\pm$ are the superpartner potentials,
\begin{equation}\label{potential1}
V_{\pm}=W^{2}\pm\frac{dW}{d\hat{r}_{*}}\,,
\end{equation}
written in terms of the so-called superpotential $W$,
\begin{equation}\label{superpotential}
W=\frac{i\sqrt{a(r)\Delta}\,(\hat{m}^{2}+{\mu_{s}}^{2}
  r^{2})^{3/2}}{r^{2}(\hat{m}^{2}+\mu_{s}^{2} r^{2})+\frac{\mu_{s} \hat{m}\sqrt{a(r)}\Delta}{2\omega}}\,.
\end{equation}
In the case of a massless spinor ($\mu_{s}=0$) the superpartner potentials reduce to 
\begin{equation}\label{potential-weyl}
V_{\pm}=\left(-\frac{m^{2}M}{l^{4}}\mp\frac{mM}{l^{5}}\sqrt{r^{2}-Ml^{2}}\right)r^{2}
+\left(\frac{m^{2}}{l^{6}}\pm\frac{2m}{l^{7}}\sqrt{r^{2}-Ml^{2}}\right)r^{4}\,.
\end{equation}

\section{Numerical results}

Although in the scalar case we already found the corresponding QNF, the numerical analysis is an interesting tool to verify the applicability of certain numerical methods in asymptotically Lifshitz spacetimes. Specifically, here we use the Finite Difference and the Horowitz-Hubeny~\cite{hubeny} methods.
Let us begin with the scalar perturbation. 
By rewriting Eq.(\ref{sc-pot}) in terms of a new variable $z=r^2$, we obtain
\begin{equation}\label{pot1}
V(r) = \frac{z}{l^8} \left[ \left( \frac{7}{4} + m^2 l^2 \right) z^2 - \left( \frac{5}{2} + m^2 l^2 - \frac{\kappa^2l^2}{z_h} \right) z_h z + \left( \frac{3}{4} - \frac{\kappa^2l^2}{z_h} \right) z_h^2 \right]\,,
\end{equation}
where $z_h = r_h ^2$. The parable in brackets tends to infinity as long as $\left(\frac{7}{4} +m^2l^2\right)>0$, which is consistent with the Breitenlohner-Freedman-type bound for the present case. The roots of this polynomial potential are given by
\begin{eqnarray}\label{roots}
z_0 &=&0\,,\nonumber\\
z_+ &=& z_h \,, \\
z_- &=& z_h \left[ \frac{ \frac{3}{4} - \frac{\kappa^2l^2}{z_h} }{ \frac{7}{4} + m^2l^2 } \right]\,.
\end{eqnarray}
If $m^2l^2>-1$, we see that $z_-<z_+$. Thus, going back to the
original variable $r$, the roots of the potential are $r=0$ with
double multiplicity, $r=\sqrt{z_-}$ and $r=r_h$ ($r=-\sqrt{z_-}$ and
$r=-r_h$ are excluded as $r>0$). Then, since $r_h$ is the biggest root
and $V(r)$ tends to $\infty$ when $r$ tends to $\infty$, the potential is positive-definite in the region $(r_h,\infty)$. Therefore, the quasinormal modes for $m^2l^2>-1$ are necessarily stable~\cite{hubeny}.

%%%%%%%%%%%%%%%%%%%%%%%%%%%%%%%%%%%%%%%%%%%%%%%%%%%%%%%%%%%%%%%%%%%%%
\begin{figure}[h]
\begin{minipage}{7.5cm}
\includegraphics[width=6cm, height=8cm,angle=270]{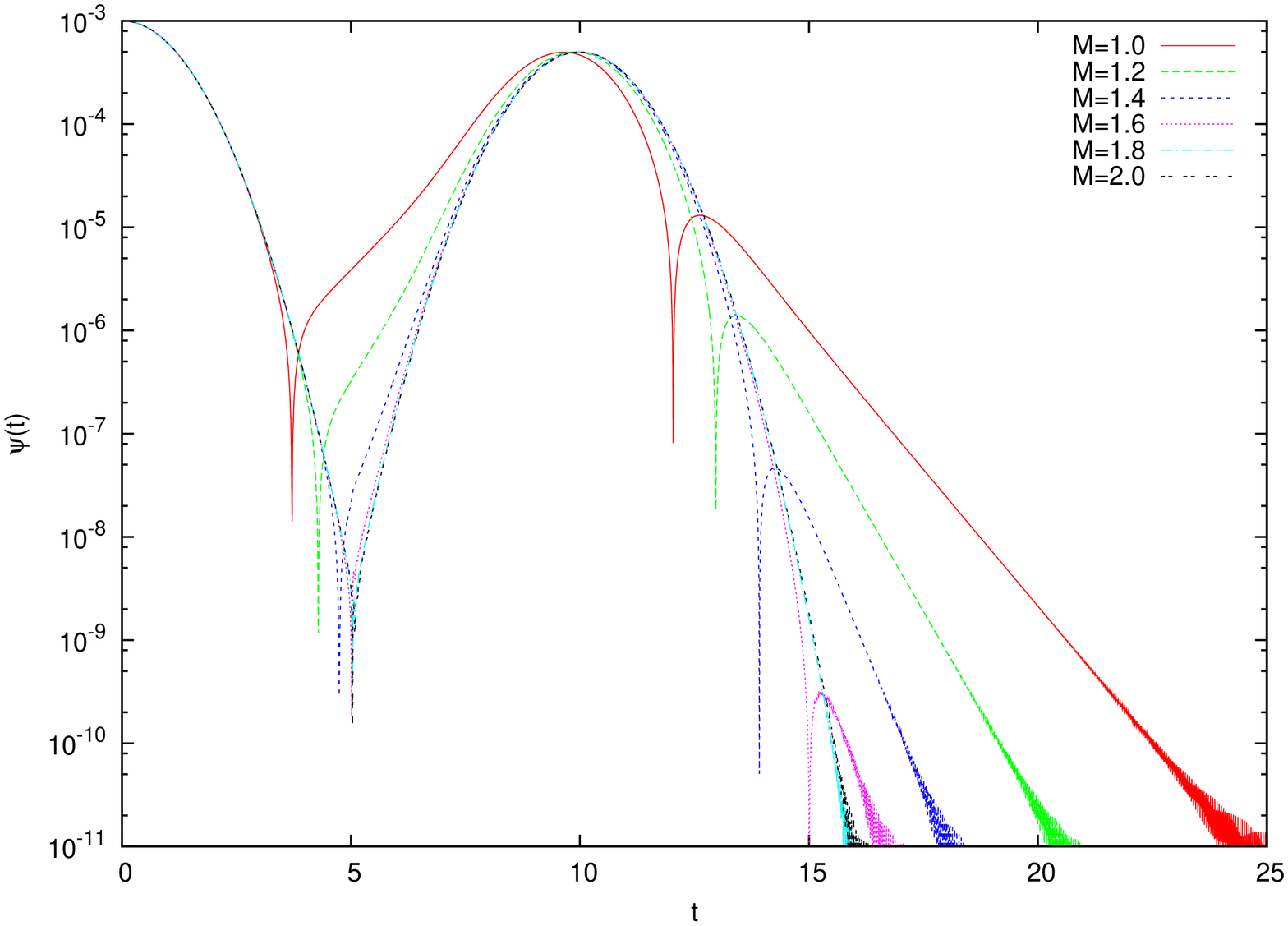}
\caption{\label{escalar_varM}Decay of scalar field with mass $m=1$ and $l=1$ for different
  values of black hole mass $M$.}
\end{minipage}\hspace{0.8cm}%
\begin{minipage}{7.5cm}
\includegraphics[width=6cm, height=8cm,angle=270]{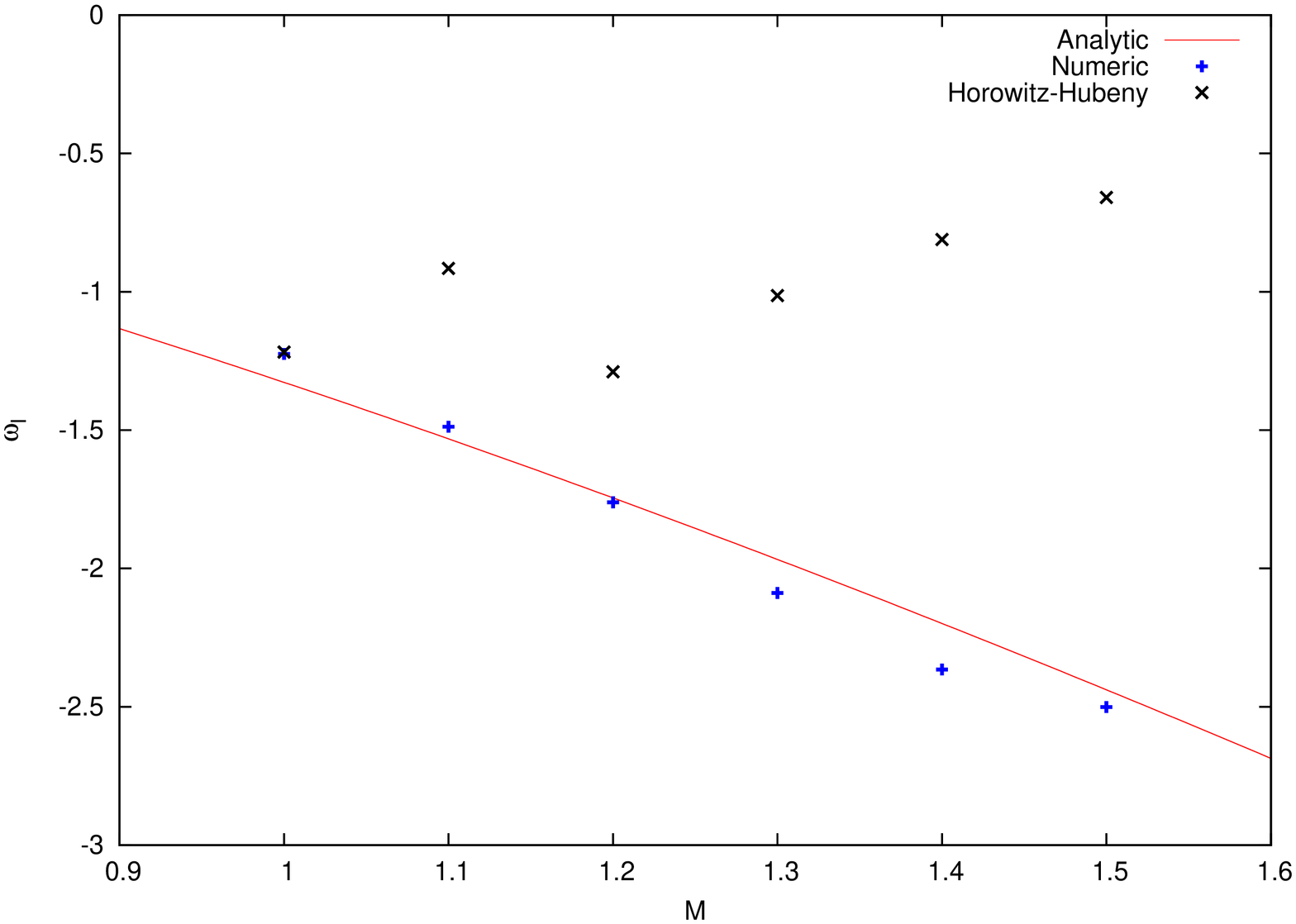}
\caption{\label{comparacao}Imaginary part of scalar quasinormal frequencies. We display the results obtained using different approaches.}
\end{minipage} 
\end{figure}
%%%%%%%%%%%%%%%%%%%%%%%%%%%%%%%%%%%%%%%%%%%%%%%%%%%%%%%%%%%%%%%%%%%%%

In Figs.\ref{escalar_varM}-\ref{comparacao} we show the numerical results for the decay of the scalar field and a comparison of the scalar QNF obtained by three different approaches, analytical calculation, finite difference, and Horowitz-Hubeny methods. Our results reinforce the conclusion already
found analytically, the $z=3$ Lifshitz black hole is stable under
scalar perturbations. Moreover, according to Fig.\ref{comparacao}, the
numerical results have a very good agreement with the analytical
calculation. However, the Horowitz-Hubeny method gives unreliable
results. We discarded explanations related to the not-convergence of the 
frequencies or the size of the black holes ocurring in other 
cases~\cite{carlos,jaqueline}, in spite of that, there is no clear explanation 
for the limitation of the method. In our case the asymptotic behavior of 
the black hole under study might play an important role in the convergence 
of the method. Nonetheless, a general criteria for the convergence of the 
Horowitz-Hubeny method remains an open question.

%%%%%%%%%%%%%%%%%%%%%%%%%%%%%%%%%%%%%%%%%%%%%%%%%%%%%%%%%%%%%%%%%%%

\begin{figure}[h]
\begin{minipage}{7.5cm}
\includegraphics[width=6cm, height=8cm,angle=270]{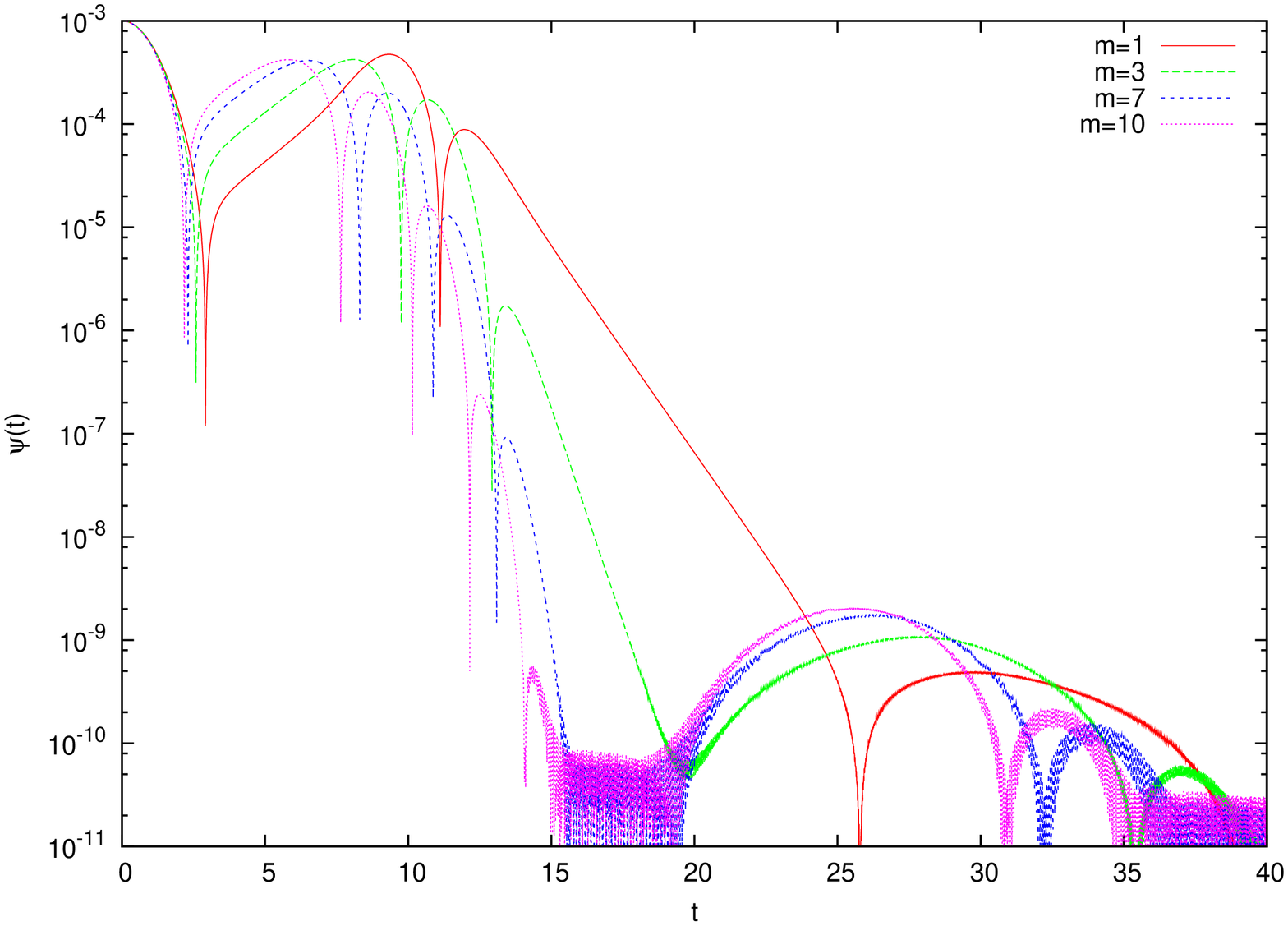}
\caption{\label{espinorM10}Decay of massless spinor with $l=1$ and black hole mass $M=1.0$ for different values of the azimuthal parameter $m$.}
\end{minipage}\hspace{0.8cm}%
\begin{minipage}{7.5cm}
\includegraphics[width=6cm, height=8cm,angle=270]{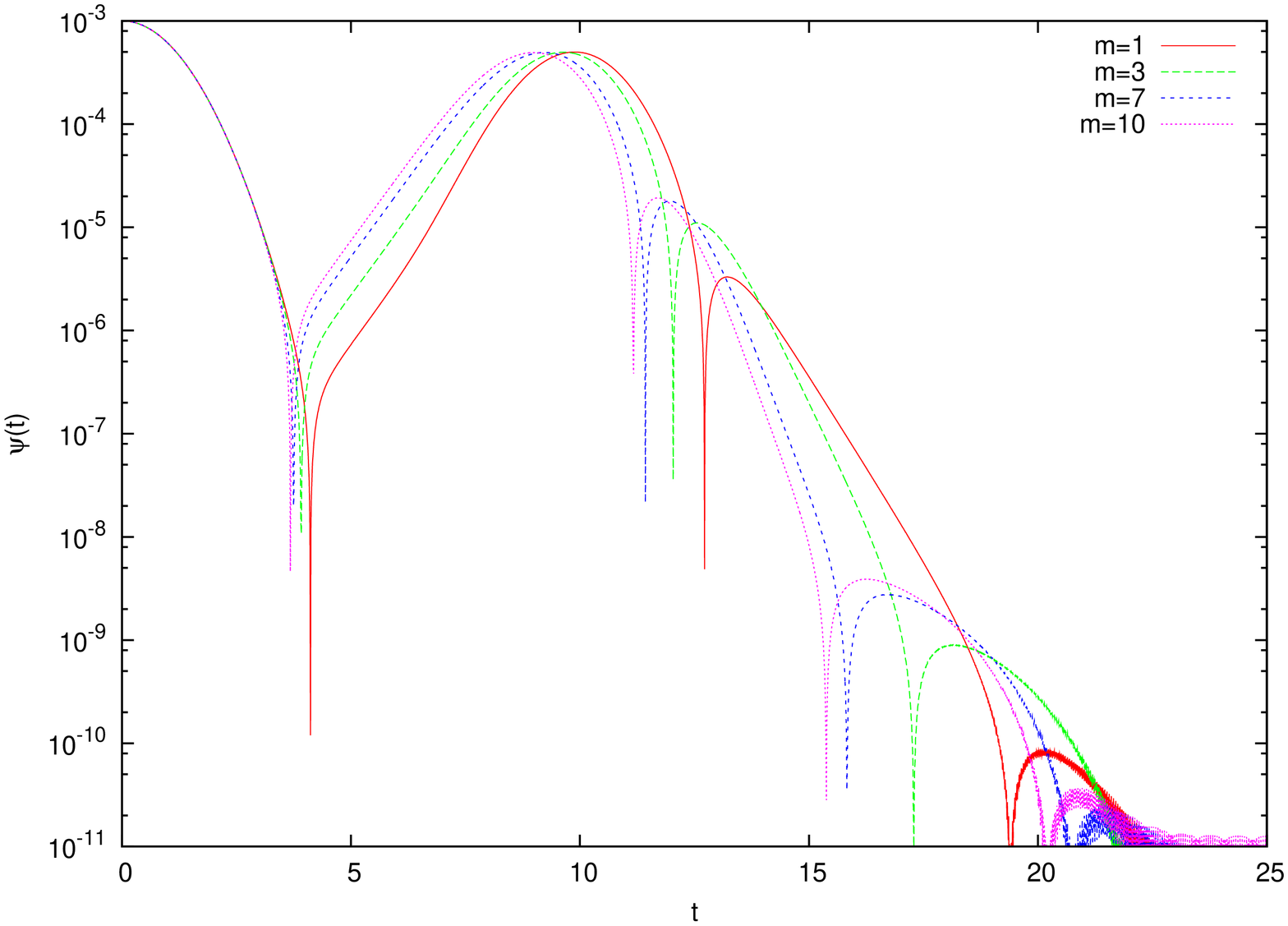}
\caption{\label{espinorM15}Decay of massless spinor with $l=1$ and black hole mass $M=1.5$ for different values of the azimuthal parameter $m$.}
\end{minipage} 
\end{figure}

%%%%%%%%%%%%%%%%%%%%%%%%%%%%%%%%%%%%%%%%%%%%%%%%%%%%%%%%%%%%%%%%%%%

Now turning to the case of spinorial perturbations, we can analogously analyze 
the superpartner potentials (\ref{potential-weyl}) and its derivatives, which we rewrite as
\begin{equation}
V_{\pm}=\frac{1}{l^8}\left[(ml)^2 r^2 (r^2 -r_+ ^2) \pm (ml)r^2 (2r^2-r_+ ^2)\sqrt{r^2-r_+ ^2}\right]\,, 
\end{equation}
\begin{equation}
V_\pm ' =  \frac{1}{l^8}\left\{ (ml)^2 r (2r^2 -r_+ ^2) \pm (ml)\left[2r(r^2-r_+ ^2)\sqrt{r^2-r_+ ^2} +r^3 \frac{2r^2-r_+ ^2}{\sqrt{r^2-r_+ ^2}}\right]\right\} \,.
\end{equation}
We see that outside the event horizon $V_+$ is positive-definite
if $ml>0$, and $\lim_{r\rightarrow\infty} V_+(r)=-\infty$ if
$ml<0$. Whereas $V_-$ is positive-definite if $ml<0$, and
$\lim_{r\rightarrow\infty} V_-(r)=-\infty$ if $ml>0$. Furthermore, we
notice that if $ml=0$, we have a free-particle case. The decaying behavior 
of the massless spinor is given in Figs.\ref{espinorM10}-\ref{espinorM15}. 
Thus, we conclude that the $z=3$ Lifshitz black hole is stable under 
massless spinorial perturbations.

\section{Applying QNM to thermodynamics}

In this section we use our analytic results for the QNF to find the area spectrum of the Lifshitz black hole. In order to apply Maggiore's method we need the asymptotic value of the QNF given in (\ref{asymptotic}), so that the proper physical frequency of the damped harmonic oscillator equivalent to the black hole QNM turns out to be
\begin{equation}\label{physfreq}
\omega_p = 2(\sqrt{6}-2) \frac{M^{3/2}}{l} N\,.
\end{equation}
Now we can calculate the adiabatic invariant
\begin{equation}
I =\int \frac{d{\cal M}}{\Delta\omega} = \int\frac{M}{\Delta\omega}dM\,,
\end{equation}
where ${\cal M}=M^2/2$ is the ADM mass of the Lifshitz black hole, and $\Delta\omega$ is the change of proper frequency between two neighboring modes, which can be obtained from (\ref{physfreq}). Thus,
\begin{equation}\label{invariant}
I = \frac{l M^{1/2}}{(\sqrt{6}-2)}\,,
\end{equation}
which is quantized via Bohr-Sommerfeld quantization in the semiclassical limit.  Then, expressing this equation in terms of the horizon area $A=2\pi r_+$, with $r_+=l\sqrt{M}$, we finally obtain the spectrum we look for,
\begin{equation}\label{areasp}
A = 2\pi (\sqrt{6}-2) n \hbar\,,
\end{equation}
with $n$ an integer number. As we can see, the horizon area of the Lifshitz black hole is quantized and equally spaced, a result not expected from a theory containing higher order curvature corrections. In addition, since this black hole has an entropy proportional to its horizon area~\cite{myung,entrop}, our result (\ref{areasp}) also says that the entropy is quantized with an evenly spaced spectrum. Subsequent studies on area spectrum of other kind of Lifshitz black holes~\cite{luna} seem to confirm this last conclusion.

\section{Conclusions}

We have studied the stability of the three-dimensional Lifshitz black hole
under scalar and spinorial perturbations in the probe limit through the
computation of 
quasinormal modes. In addition, we have found the event horizon area
quantization as an application of the results for quasinormal modes
using Maggiore's prescription. 

Regarding the scalar perturbations, we found that the QNF are purely imaginary and negative pointing out a stable equilibrium configuration for the Lifshitz black hole. Our numerical results perfectly agree with the analytical calculation in the case of the finite difference method, which allows to obtain the temporal behavior of the fields showing all the stages of the decay. However, in the case of the Horowitz-Hubeny method the frequencies values fail to be obtained. An important observation is that apart from the numerical factor, the asymptotic scalar frequency is the same as that calculated in the hydrodynamic limit in the context of Gauge/Gravity duality~\cite{abdallajef}.

Concerning the spinorial perturbation, our numerical results also show probe massless spinors decaying, and therefore, the $z=3$ Lifshitz black hole is 
stable under spinorial perturbations too.

Additionally, using the asymptotical form of the analytical QNF we calculated the area spectrum of the Lifshitz black hole using Maggiore's method. Our results show that the horizon area and the entropy are quantized and their spectra are equally spaced.

At last, we should stress that the definite answer on the stability of the Lifshitz black hole should come from gravitational perturbations, mainly from the tensor part of the metric perturbation, since NMG admits the propagation of gravitational waves contrary to Einstein gravity in three dimensions, that has no propagating degrees of freedom.

\ack{
This work was supported by Funda\c c\~ao de Amparo \`a Pesquisa do Estado de S\~ao Paulo {\bf (FAPESP-Brazil)} and Consejo Nacional de Investigaciones Cient\'{i}ficas y T\'ecnicas {\bf (CONICET-Argentina)}.}

\end{document}